\documentclass[preprint,prl,aps] {revtex4}
\usepackage{epsfig}
\newcommand{\beq}{\begin{equation}}
\newcommand{\eeq}{\end{equation}}
\usepackage{graphics,rotating}
\usepackage[latin1]{inputenc}
\begin{document}
\author{Zilvinas Rinkevicius}
\affiliation{Laboratory of Theoretical Chemistry,
The Royal Institute of Technology, SCFAB, SE-10691 Stockholm, Sweden}
\author{Gennady P. Berman}
\affiliation{Theoretical Division,T-13, MS B213, Los Alamos National
  Laboratory, Los Alamos, New Mexico 87545}
\author{David L. Allara}
\affiliation{Pennsylvania State University, University Park,
Pennsylvania 16802}
\author{Vladimir I. Tsifrinovich}
\affiliation{IDS Department, Polytechnic University, Six Metrotech
Center, Brooklyn, New York 11201}
\author{Sergei Tretiak}
\affiliation{Theoretical Division,T-12, MS B268, Los Alamos
National Laboratory, Los Alamos, New Mexico 87545}

\title{Characteristic parameters and dynamics of two-qubit
system\\ in self-assembled monolayers}

\begin{abstract}
We suggest the application of nitronylnitroxide substituted with
methyl group and  2,2,6,6-tetramethylpiperidin organic radicals as
$1/2$-spin qubits for self-assembled monolayer quantum devices. We
show that the oscillating cantilever driven adiabatic reversals
technique can provide the read-out of the spin states. We compute
components of the $g$-tensor and dipole-dipole interaction tensor
for these radicals. We show that the delocalization of the spin in
the radical may significantly influence the dipole-dipole
interaction between the spins.
\end{abstract}

\maketitle

\section{1. Introduction}

The self assembled monolayer (SAM) molecular systems with no or
small amount of defects are the promising candidates for many
electronic devices \cite{SCIENCE}. SAM systems containing radicals
with unpaired spin $1/2$ can be used in quantum logic devices
\cite{berm02}. It was suggested using micro-wires, which produce a
magnetic field gradient, and {\it rf} pulses, which induce Rabi
transitions, to manipulate with the radical spins and create the
quantum entanglement between them. Later the conditions for the
molecules appropriate for the quantum logic devices have been
formulated, and 1,3-diketone radicals have been suggested and
theoretically analyzed from the point of view of their application
to the quantum logical devices \cite{tamulis}.

This paper pursues two objectives. First, we suggest two stable
organic radicals, nitronylnitroxide substituted with methyl group
(NITRO) and 2,2,6,6-tetramethylpiperidin (TEMPO), for SAM quantum
devices. In order to stimulate the experiments we compute the
$g$-tensor and zz-component of the dipole-dipole interaction
tensor. Second, we suggest the novel oscillating cantilever driven
adiabatic reversals (OSCAR) technique as the tool for read-out of
the spin states. The OSCAR technique has been recently
successfully implemented for the single-spin detection
\cite{natu04}. In section II we describe the effective spin
Hamiltonian for the two interacting radicals, in section III we
consider the method of creation and detection entanglement between
the two spins, in section III we describe the methods of
computation of the $g$-tensor and the dipole-dipole interaction
tensors, and in section IV discuss the results of our quantum
chemical computations.

\section{2. Spin Hamiltonian for two qubits}

Let us discuss the simplest possible quantum computer element -
the two qubit system. In SAM the two organic radicals are embedded
in a monolayer.  Electron spins $S=1/2$ localized on the radicals
represent the two qubit system. We use geometrical arrangement the
same as in the Ref. [\onlinecite{berm02}], where external magnetic
field $B_0$ is oriented along the $z$-axis and radicals are
separated by the distance $a$ along the $y$-axis (see Fig. 1). In
addition to constant external magnetic field $ B_0$ a gradient of
external magnetic field $\partial B_z/\partial y$ along the
$y$-axis is applied on the SAM. If radicals in SAM are oriented in
such a way that the principle $z$-axis of the electronic g-tensor
corresponds to the orientation of the external magnetic field
${\bf B}_0$, then the effective Hamiltonian for the spin group is
\begin{equation}
H_{spin} = \mu_B g_{zz} B_0 S_z(2)  + \mu_B g_{zz}
(B_0-\frac{\partial B_z}{\partial
  y}a)S_z(1) - D_{zz}S_z(1)S_z(2) \ ,
\label{Hspin}
\end{equation}
where $S_z(1)$, $S_z(2)$  are respectively radical one and radical
two spin projection operators, $\mu_B$ - the Bohr magneton,
$g_{zz}$ - the component of the electronic g-tensor of a radical
(we assume that the radicals are the same in the spin group) and
$D_{zz}$ is the dipole-dipole interaction tensor component.  The
Hamiltonian (\ref{Hspin}) can be treated in electron spin
functions basis $\vert \sigma_1 \sigma_2 \rangle$ ($\sigma_1$,
$\sigma_2$ = 0 or 1, where 0 stands for the spin ground state and
1 for the spin excited state). As in our system the magnetic field
gradient along the $y$-axis exists, both radicals in the spin
group (first and second) have distinguished spin transition
frequencies, which depend on the local magnetic field strength at
the radical position and additionally shifted by the dipole-dipole
interaction between the radicals.

Under these conditions, energy levels of the system are defined by
the following expressions:
\begin{equation}
\vert 0 0 \rangle:~  E = -\frac{1}{2}\mu_B g_{zz} B_0 -
\frac{1}{2}\mu_B g_{zz} (B_0 +Ga) +\frac{1}{4} D,
\end{equation}
$$
\vert 0 1 \rangle:~  E =  \frac{1}{2}\mu_B g_{zz} B_0 -
\frac{1}{2}\mu_B g_{zz} (B_0 +Ga) - \frac{1}{4} D,
$$
$$
\vert 1 0 \rangle:~  E = -\frac{1}{2}\mu_B g_{zz} B_0 +
\frac{1}{2}\mu_B g_{zz} (B_0 +Ga) - \frac{1}{4} D,
$$
$$
\vert 1 1 \rangle:~  E = \frac{1}{2}\mu_B g_{zz} B_0 +
\frac{1}{2}\mu_B g_{zz} (B_0 +Ga) + \frac{1}{4} D,
$$
$$
G=|\partial{B_z}/\partial y|,~D=|D_{zz}|,
$$
($D_{zz}$ and $\partial{B_z}/\partial y$ are negative), the
external magnetic field on the second spin is $B_0$ and on the
first spin is $B_0+Ga$). The energy levels diagram of two qubit
system with four possible transitions is given in Fig.~2. The
transition frequencies are determined by the following expressions
($\hbar=1$) :
\begin{equation}
\vert 00 \rangle - \vert 01 \rangle:~ \omega^0_2 = \mu_B g_{zz}B_0
- \frac{1}{2}D,
\end{equation}
$$
\vert 00 \rangle - \vert 10 \rangle:~ \omega^0_1 = \mu_B
g_{zz}(B_0 +Ga) - \frac{1}{2}D,
$$
$$
\vert 10 \rangle - \vert 11 \rangle:~ \omega^1_2 = \mu_B g_{zz}B_0
+ \frac{1}{2}D,
$$
$$
\vert 01 \rangle - \vert 11 \rangle:~ \omega^1_1 = \mu_B
g_{zz}(B_0 +Ga)+ \frac{1}{2}D.
$$
Here $\omega^k_i$ means the transition frequency for the spin
``i'' ($i=1,2$) if the neighboring spin is in the state
$|k\rangle$ ($k=0,1$).  Given formulas for radical spin energy
levels involve two molecular parameters: electronic g-tensor
component $g_{zz}$ (property of radical itself) and dipole-dipole
coupling parameter $D$, which is the property of the two radical
system. Both these properties can be efficiently evaluated using
quantum chemistry methods and influence of various parameters
(radicals geometries, distance between radicals in spin group,
radical relative orientation to each other, etc.) on these
properties can be investigated.

\section{3. Entangled spin states}

In this section, we consider the creation and detection of the
entangled state
$$
{{1}\over{\sqrt{2}}}[|00\rangle+\exp(i\phi)|11\rangle],
$$
where $\phi$ is a nonsignificant phase factor.

First, one applies a $\pi/2$-pulse with the frequency $\omega^0_1$
(we assume that initially both spins are in their ground state
$|00\rangle$, as it is shown in Fig. 1). This pulse drives the
spin system into the state
$$
{{1}\over{\sqrt{2}}}[|0\rangle+\exp(i\phi_1)|1\rangle]|0\rangle.
$$
Now one applies a $\pi$-pulse with the frequency $\omega_2^1$.
This pulse drives the second spin if the first spin is in the
state $|1\rangle$. Thus, the new state of the system is
$$
{{1}\over{\sqrt{2}}}[|00\rangle+\exp(i\phi_2)|11\rangle].
$$
This is a desired entangled state of the two-qubit system.

To detect this state one uses the OSCAR technique. In the OSCAR
technique the cantilever tip with the ferromagnetic particle
oscillates along the line which is parallel to the sample surface.
The distance between the ferromagnetic particle and the selected
radical spin changes periodically.  When the cantilever tip is in
its end point one starts to apply the {\it rf} field. The magnetic
field on the spin changes its magnitude with the cantilever
period. Suppose that the equilibrium position of the cantilever
corresponds to the resonant condition for the spin. Then, in the
rotating system of coordinates the effective field on the spin
changes its direction from +z to -z in the x-z plane (we assume
that the rotating {\it rf} field points in the positive
x-direction of the rotating system). The condition of the
adiabatic reversals is
\begin{equation}
\vert  \frac{d {\vec B}_{ef}}{dt} \vert  << \mu_B g_e B_1^2,
\end{equation}
where ${\vec B}_{ef}$ is the effective magnetic field in the
rotating frame, $B_1$ is the {\it rf} field amplitude, and
$g_e$ is the free electron g-factor.
If the condition of the adiabatic reversals is satisfied the spin
follows the effective field.

The back action of the spin on the cantilever tip causes the
frequency shift of the cantilever vibrations, which can be
measured with high accuracy. The cantilever frequency shift can
take two values $\pm|\delta\omega_c|$ depending on the spin
direction relatively the effective field. Measuring the sign of
the frequency shift one can find the initial direction of the spin
relatively to the effective magnetic field. We assume that
initially (when we start application of the {\it rf} field) the
effective magnetic field has the same direction as the external
magnetic field. In this case the ground state of the resonant spin
corresponds to the negative frequency shift of the cantilever
vibrations \cite{berm03}.

How to verify the entangled state of the two-spin system? The
entangled state manifests itself in the two outcomes of the
measurement: $\vert 00\rangle$ or $\vert 11 \rangle$.  Let, for
example, we apply the {\it rf} field with the frequency $\omega =
\omega_2^0$. In general, we have three possible outcomes for the
measurement of the cantilever frequency shift: 1) The cantilever
frequency shift is negative. It means that the second (right) spin
is in the ground state, and the first (left) spin is also in the
ground state. Thus, our system collapsed to the state $\vert 00
\rangle$ (in the system of coordinates connected to the effective
field). 2)
 The frequency shift is zero. It means that our system initially collapsed to
the states $\vert 11 \rangle$ or $\vert 10 \rangle$ (or their
superposition), which are insensitive to the frequency $\omega =
\omega_2^0$. In this case, after the integer number of the
cantilever cycles we may change the {\it rf} field frequency from
$\omega = \omega_2^0$ to, for example, $\omega = \omega_1^0$. Now
we have two possible outcomes: 2a) For the state $\vert 10\rangle$
the left resonant spin will provide the positive cantilever
frequency shift; 2b) For the state $\vert 11 \rangle$ there is no
resonant spin, and the frequency shift is zero. 3) The frequency
shift is positive. It means the second resonant spin is in the
excited state $\vert 1 \rangle$, and the first spin is in the
ground state $\vert 0\rangle$, i.e. our system has collapsed to
$\vert 01 \rangle$. Thus, the repeated outcomes 1) or 2b) (with
equal probability)  correspond to the initial entangled state,
which collapses to the states $\vert 00 \rangle$ or $\vert 11
\rangle$. The outcomes 2a) or 3) will show that the system is not
in the expected entangled state.  One can see that the OSCAR
technique provides the simple verification of the quantum
entanglement. Note that the verification experiment must be
conducted for the time interval smaller than the characteristic
time between the spin quantum jumps.

\section{4. Computational methods}

{\bf {A. Electronic $g$-tensor}}

According to Spin Hamiltonian (see Eq.~\ref{Hspin}) electronic
$g$-tensor of a radical is defined as the second derivative of the
molecular electronic energy $E$
\begin{equation}
{\bf g} = \frac{1}{\mu_B}\frac{\partial^2E}{\partial {\bf B}\partial{\bf S}}
\bigg \vert_{{\bf B}={\bf 0},{\bf S}={\bf 0}} \ .
\label{eq:g_def}
\end{equation}
For molecules described by a Breit-Pauli Hamiltonian, where the
spin and the magnetic field as well as all other relativistic
corrections are treated in perturbation theory framework, the
molecular electronic $g$-tensor can be evaluated using expression
(up to second order in the fine structure constant $\alpha$)
\cite{rink03}:
\begin{equation}
{\bf g} = g_e{\bf 1} + \Delta{\bf g}_{\rm RMC} + \Delta{\bf g}_{\rm GC(1e)} +
\Delta{\bf g}_{\rm GC(2e)} + \Delta{\bf g}_{\rm OZ/SO(1e)} +
\Delta{\bf g}_{\rm OZ/SO(2e)} \ .
\label{eq:gten}
\end{equation}
In this equation, the first terms is the free electron g-factor,
which comes from the electronic Zeeman  operator in the
Breit-Pauli Hamiltonian, with the radiative corrections from
quantum electrodynamics introduced into the Hamiltonian
empirically. The next three terms originate from first order
perturbation theory applied to the Breit-Pauli Hamiltonian and are
the mass-velocity and the one- and two electron corrections to the
electronic Zeeman effect. The last two terms are, respectively,
the one- and two-electron corrections contributing to the
electronic g-tensor  to second order in perturbation theory as
cross terms between the spin-orbit operators and the orbital
Zeeman operator. All terms except the free electron g-factor
value, $g_e$,  in Eq.~(\ref{eq:gten}) contribute to the electronic
g-tensor shift $\Delta {\bf g}$, which accounts for the influence
of the local electronic environment in the molecule on the
unpaired electrons. From the contributions to the g-tensor shift
given in Eq.~(\ref{eq:gten}), the first three terms are evaluated
straightforwardly according to Eq.~(\ref{eq:g_def}) from
expectation values of the corresponding Breit-Pauli Hamiltonian
operators.  The last two terms, namely spin--orbit (SO)
corrections, are considerably more involving computationally, as
they are defined in terms of second-order perturbation theory,
{\it
  i.e.\/} their calculation formally require a knowledge of all
relevant excited states energies and wave functions. The
evaluation of these terms using density functional theory  (DFT)
methods are most often done using various kinds of  approximate
sum-over-states approaches or response theory methods
\cite{malk02,nees01,kaup02,rink03}.

In the following part of this section we briefly describe electronic
$g$-tensor evaluation approach implemented in quantum chemistry code
DALTON \cite{dalton}, which is based on a restricted DFT linear response
theory \cite{rink03}. In this approach from the first-order
perturbation theory contributions to the electronic
g-tensor shift, we only include those terms that involve
one-electron operators as DFT in principle can not handle two-electron
operators \cite{rink03}. The Cartesian $ab$ components of these
contributions to $\Delta {\bf g}$ tensor are
\begin{equation}
\Delta{\bf g}_{\rm RMC}^{ab}= - \frac{\alpha^2}{S}\langle 0 \vert
 \sum_i p^2_i s_{z,i} \vert 0 \rangle \delta^{ab},
 \end{equation}
 $$
\Delta{\bf g}_{\rm GC(1e)}^{ab}= \frac{\alpha^2}{4S} \langle 0
\vert \sum_i\sum_N \frac{Z_N}{r_{iN}^3}[({\bf r}_{iN}\cdot{\bf
r}_{iO})\delta^{ab}- {\bf r}_{iO}^a\cdot{\bf
r}_{iN}^b]s_{z,i}\vert 0 \rangle,
$$
where $p_i$ is the canonical linear momentum of electron $i$, $s_{z,i}$ the
z-component of the spin operator of electron $i$, ${\bf r}_{iN}$
 and
${\bf r}_{iO}$ are the position vectors of electron $i$ relative
to nucleus $N$ and the magnetic gauge origin $O$, respectively. In
the above equations $\Psi_0$ is chosen to be the ground-state wave
function with maximum spin projection $S=M_S$. Neglecting of the
two-electron gauge correction, $\Delta{\bf g}_{\rm GC(2e)}$, the
contribution to the $g$-tensor shift in DFT calculations does not
influence the accuracy of the g-tensor evaluation as this term
only gives a correction from a two-electron screening of the
$\Delta{\bf g}_{\rm GC(1e)}$  and considering the smallness of the
one-electron gauge correction term itself is justified
\cite{malk02,kaup02,rink03}. The major contributions to  $\Delta
{\bf g}$ tensor arising from second-order perturbation theory,
so-called one- and two-electron SO corrections, are evaluated as
linear response functions
\begin{equation}
\Delta{\bf g}_{\rm OZ/SO(1e)}^{ab}= \frac{1}{S} \langle \langle
l_{iO}^a; H_{\rm SO(1e)}^b \rangle \rangle_0,
\end{equation}
$$
\Delta{\bf g}_{\rm OZ/SO(2e)}^{ab}= \frac{1}{S} \langle \langle
l_{iO}^a; H_{\rm SO(2e)}^b \rangle \rangle_0,
$$
where the spectral representation of the linear response function at zero
frequency for two arbitrary operators $\hat H_1$ and $\hat H_2$
is given by
\begin{equation}
\langle\langle \hat H_1; \hat H_2 \rangle\rangle_0 = \sum_{m>0} \frac {\langle
0| H_1 |m \rangle\langle m| H_2 | 0 \rangle + \langle 0| H_2
|m \rangle\langle m| H_1 | 0 \rangle} {E_0-E_m} \ .
\label{sum}
\end{equation}
In Eq. (8) $l_{iO}^a$ is the cartesian $a$ component of the
angular momentum operator of electron $i$, and $H_{\rm SO(1e)}^b$
and $H_{\rm SO(2e)}^b$ are the Cartesian $b$ components of the
one- and two-electron SO operators. The Cartesian component $b$ of
the one-electron SO operator is defined as
\begin{eqnarray}
H_{\rm SO(1e)}^b &=&   \frac{\alpha^2}{2} \sum_i \sum_N \frac{Z_N l^b_{iN}}{r_{iN}^3}s_{
z,i}, \label{eq:HSO1e}
\end{eqnarray}
As mentioned above in DFT two-electron operators can not be
evaluated properly and one need to introduce one or another
approximation of the two-electron operators in order to perform
calculations within limits of formalism. For g-tensor calculations
performed in this work we selected to approximate two-electron SO
operators by an  Atomic Mean Field (AMFI) SO operator~\cite{amfi}
as previous experience with the AMFI SO approximation  in {\em ab
intio} and DFT works devoted to  electronic g-tensors calculations
have been very encouraging and no significant problems with the
accuracy of the AMFI SO approximation has been reported
\cite{malk02}.

{\bf {B. Dipole-dipole coupling}}

 Formally, in the case of two radicals with the spins $S=1/2$, the
interaction operator and corresponding ${\bf D}$ tensor may be
written as
\begin{equation}
H_{DD} = \alpha^2 \mu_B^2 g_e^2 \frac{3({\bf S}_1 \cdot {\bf
r})({\bf S}_2 \cdot {\bf r}) -
  r^2 {\bf S}_1 \cdot {\bf S}_2}{r^5},
  \end{equation}
  $$
D_{ii} = \alpha^2 \mu_B^2 g_e^2 \langle \frac{3r_{i}^2 - r^2}{r^5}
\rangle,
$$
$$
D_{ij} = \alpha^2 \mu_B^2 g_e^2 \langle
\frac{3r_{i}r_{j}}{r^5}\rangle,~ ({i,j=x,y,z}),~(i\not= j),
$$
where ${\bf S}_1$, ${\bf S}_2$ are respectively radical one and
radical two effective spin operators, ${\bf r}$ - the ${\bf S}_2$
position vector with respect to ${\bf S}_1$. In above definition
of the dipole-dipole interaction tensor we assumed that each
radical electronic g-tensor is isotropic and equal to the free
electron g-factor. In this work we selected two methods for
calculation of the dipole-dipole interaction tensor required for
determination of transition frequencies in the two qubit system
(see Eqs. (3)): classical point dipole approach and ``point
dipole--spin density'' approach. In case of classical point-dipole
approximation delocalization of unpaired electrons is neglected
and {\bf D} tensor components are evaluated (in our two spin
system geometry, {\bf r}=(0,a,0)) as
\begin{equation}
D_{zz} = D_{xx} = - \alpha^2 \mu_B^2 g_e^2 \frac{1}{a^3},
\end{equation}
$$
D_{yy} = 2 \alpha^2 \mu_B^2 g_e^2 \frac{1}{a^3},
$$
$$
D_{ij}=0,~ (i\not= j).
$$
One can expect this approach to be accurate for large separation
of radicals, then the unpaired electrons delocalization region are
very small comparing to the distance between the radicals.
However, in case of moderate separation between radicals (0.5-1.5
nm) unpaired electrons delocalization in radicals can not be
neglected and should be explicitly taken into account evaluating
spin-dipole interaction tensor {\bf D}. The simplest approach,
which partially accounts for unpaired electrons distribution in
radicals, is ``point dipole--spin density'' approach in which one
radical unpaired electron treated as delocalized and another
radical electron magnetic moment represented as a point dipole. In
this approximation {\bf D} tensor components are evaluated as
\begin{equation}
D_{ii} = \alpha^2 \mu_B^2 g_e^2 \int \rho({\bf r}_e)
\frac{3r_{i}^2 - r^2}{r^5} d V,
\end{equation}
$$
D_{ij} = \alpha^2 \mu_B^2 g_e^2 \int \rho({\bf r}_e)
\frac{3r_{i}r_{j}}{r^5} d V,~({i,j=x,y,z}),~(i\not= j),
$$
where $\rho({\bf r}_e)$ is the electron spin density distribution
in radical and ${\bf r}$ is the electron position with respect to
the magnetic dipole position ${\bf r}_M$, ${\bf r} = {\bf r}_e
-{\bf r}_M$. The described approach only partially takes into
account delocalization of the unpaired electrons in ${\bf D}$
tensor calculations, but in order to obtain qualitative picture of
the electron delocalization influence on the dipole-dipole
interaction of the two radicals is sufficient.  In this work we
use this approach to determine limits of the classical point
dipole approach for evaluation of the dipole-dipole interaction
tensor ${\bf D}$ in two radicals system.

\section{5. Computational details}

As the promising candidates for qubits in SAM of organic radicals
we selected two well-known stable organic radicals, namely methyl
group substituted nitronyl nitroxide (NITRO) and
2,2,6,6-tetramethylpiperidin (TEMPO) (see Fig. 3), which form
stable organic crystals and already have applications in material
sciences as well as biochemistry. Both chosen radicals possess one
unpaired electron and therefore each radical can be recognized as
a single qubit. The geometry of NITRO and TEMPO radicals used in
calculations of electronic g-tensors and of dipole-dipole
interaction tensor {\bf D} have been obtained by performing
geometry optimization of single radical in 6-311G(d,p) basis set
\cite{6311g} at the B3LYP \cite{becke,lyp,b3hyb} level . All
geometry optimizations have been carried out in Gaussian-98
program \cite{gaus98}. Apart from optimizing geometries of NITRO
and TEMPO radicals we performed geometry optimization of the NITRO
and TEMPO radicals derivatives with ${\rm (CH_2)_n}$ tails (see
Fig.~4). The ${\rm (CH_2)_n}$ groups were added to radicals in
order to simulate conventional structure of the compounds used in
formation of SAM, which usually feature long ${\rm (CH_2)}$ groups
or similar tails. Obtained structures of NITRO and TEMPO radicals
with ${\rm (CH_2)_n}$ tails have been used to investigate
influence of the tails on the properties of the NITRO and TEMPO
derivatives compared to free NITRO and TEMPO radicals and allowed
us to estimate feasibility of the  NITRO and TEMPO radicals usage
as basic building blocks of the spin arrays in SAM.

Calculations of electronic g-tensors for NITRO and TEMPO radicals as well as their derivatives
with ${\rm (CH_2)}$ tails have been carried out using BP86 exchange--correlation functional
\cite{becke,pw86}, which gives accurate results for organic radicals. In all calculations of g-tensors  we
employed IGLO-II basis set \cite{iglo2} especially designed for evaluation of magnetic properties. In order to
estimate influence of the  ${\rm (CH_2)_n}$ (n=1,2) tails, which usually added to radicals in order to enable
formation of SAM, on the electronic g-tensors of the radicals we carried out electronic g-tensors
calculations for single NITRO and TEMPO radicals (see Fig.~3) as their derivatives with ${\rm (CH_2)_n}$
tails (see Fig.~4).

Calculations of the dipole-dipole interaction tensor between two
NITRO or two TEMPO radicals have been performed using previously
described point dipole and ``point dipole--spin density''
approaches varying the distance between radicals, $a$, from 1 nm
to 2 nm. Single radicals optimized geometries have been employed
in calculations using ``point dipole--spin density'' approach.
 All calculations
have been performed at the B3LYP level using Duning's double zeta
basis set \cite{dz}, which allow adequate description of the electron
density distribution in investigated radicals. In calculations of
the dipole-dipole interaction tensor we limited ourselves only by
investigation of the $D_{zz}$ component dependence on the distance
between radicals $a$ and contrary to investigation of the
electronic g-tensors do not carried out investigation of the ${\rm
(CH_2)_n}$ tails influence on dipole-dipole interaction tensor
${\bf D}$, as our g-tensor calculations indicated negligible
influence of the ${\rm (CH_2)_n}$ tails on unpaired electron
density distribution in both NITRO and TEMPO radicals.

\section{6. Results and discussion}
{\bf {A. Electronic g-tensor}}

Electronic g-tensor calculations results for single NITRO and
TEMPO molecules as well as their derivatives with  ${\rm
(CH_2)_n}$  n=1,2 tails are tabulated in Table 1.~\ref{tab:gten}
The results of electronic g-tensor calculations for NITRO compound
with ${\rm (CH_2)_n}$ tail separated for two different
conformations of this compound, which differ only by orientation
of the tail with one ${\rm (CH_2)}$ unit. However, already for
tail consisting of the two  ${\rm (CH_2)}$ units, there is no
difference between NITRO A and NITRO B conformations due to the
increased flexibility of tail, which leads geometry optimization
procedure converges to same structure independently on the
starting geometry. Here, we note even thought our calculations
predict small radical tail rotation around C-C bonds it does not
correspond  to the ``real'' behavior of the ${\rm (CH_2)}$ tail in
SAM, as the motion of the tail is constrained by surrounding
molecule tails in SAM.  Now let us turn discussion from radicals
geometrical structure features to their electronic g-tensors,
which are one of the key quantities in our two qubit system Spin
Hamiltonian.

All electronic g-tensor components, presented in Table 1,
~\ref{tab:gten} are only slightly altered by addition of ${\rm
(CH_2)_n}$ tail chain units, suggesting small distortion of
unpaired electron density in radical by such chemical
modification. The electronic g-tensor shifts almost converge, and
extension of the chain further from two to three ${\rm (CH_2)}$
units changes g-shift components in range 1-5 ppm. Therefore, we
conclude that electronic g-tensors of the NITRO and TEMPO radicals
are only slightly affected by chemical addition of the  ${\rm
(CH_2)_n}$ tail, which is required for growth of the SAM, i.e.
both radicals preserve their properties in SAM. Selected radicals
have highly anisotropic electronic g-tensor (see Table 1
~\ref{tab:gten} ) with large $\Delta g_{xx}$ and $\Delta g_{yy}$
components (for orientation of the electronic g-tensor axes see
Fig.~1). Only $\Delta g_{zz}$ is small, and therefore along this
axis we have molecular g-tensor component close to the free
electron g-factor. The large anisotropy of the electronic g-tensor
poses significant difficulties for the quantum computation
prospective as in order to have well defined frequencies of the
 transitions it is essential to have fixed
orientation of the electronic g-tensor principal axes with respect
to the external magnetic field, i.e. radicals in SAM should be
fixed in their position and rotations of the radicals with respect
to their g-tensor principle axes must be restrained.

{\bf {B. Dipole-dipole interaction}}

Dipole-dipole interaction tensor between two radicals is one of
the key parameters in the spin Hamiltonian, which describes the
two qubit system. We employed different approaches, namely
classical point dipole approach and ``point dipole--spin density''
approach, for evaluation of the $D=|D_{zz}|$.The second approach
as discussed previously partially accounts electron distribution
in molecule and therefore results are directly dependent on the
unpaired electron density localization in the molecule. Contrary,
the first approach does not account for unpaired electron
delocalization in radical and therefore gives the same results for
all radicals. The dipole-dipole interaction dependence on the
distance between radicals $a$ is plotted in Fig.~5, where results
for both classical point dipole approach (denoted classical) and
``point dipole--spin density'' approach (denoted Nitro and Tempo
for NITRO and TEMPO radicals, respectively). Quick inspection of
these plots indicates a substantial influence of the unpaired
electron delocalization on the $D$ value especially at the small
$a$ region. Therefore, unpaired electron delocalization can not be
neglected in evaluation of the ${\bf D}$ tensor and previous data,
which have been used in modeling of the qubits system, obtained
with classical point dipole approach should be carefully
reexamined for $a$ range of the 10-15 {\AA}. Another implication
of the non-negligible contribution from unpaired electron
delocalization to the dipole-dipole interaction is that
orientation of the radical can influence ${\bf D}$ tensor
components values. Therefore, similarly to electronic g-tensor
calculations the dipole-dipole interaction modeling results
suggest that the rigid fixation of the radical in SAM with
specific orientation is one of the key requirements in production
of SAM in order to make them useful in quantum computing.

\section{7. Conclusion}

We have suggested using NITRO and TEMPO radicals as spin qubits
for a quantum logical device based on SAM systems. In order to
stimulate experimental implementation of our idea we have computed
the components of g-tensor and dipole-dipole interaction tensor
for these radicals. We have shown that adding $(CH_2)_n$ tail
chain does not influence significantly the radical electron
g-factor. Delocalization of the electron spin in radicals
influences the dipole-dipole interaction between the radicals. We
suggested a scheme for detection of the entanglement between the
two radicals based  on the novel OSCAR technique, which has been
recently used for the single-spin detection.

\section*{ACKNOWLEDGEMENTS}
We are grateful to G. D. Doolen for useful discussions. This work
was supported by the Department of Energy (DOE) under Contract No.
W-7405-ENG-36, by the National Security Agency (NSA) and Advanced
Research and Development Activity (ARDA) under Army Research
Office (ARO) contract \# 707003.


\newpage
\begin{table}
\label{tab:gten} \vspace{1.0 cm} \caption{Shift of the electronic
g-tensor components on ${\rm (CH_2)_n}$
  chain in nitronyl-nitroxide and TEMPO radicals. $^{a,b}$}
\begin{tabular}{lccccccccccccccccccccccccc}
\hline \hline
 && \multicolumn{3}{c}{NITRO A} && \multicolumn{3}{c}{NITRO B} &&
 \multicolumn{3}{c}{TEMPO} \\
 \cline{3-5} \cline{7-9} \cline{11-13}
Tail && ---  & ${\rm CH_2}$ & ${\rm (CH_2)_2}$ &&
         ---  & ${\rm CH_2}$ & ${\rm (CH_2)_2}$ &&
         ---  & ${\rm CH_2}$ & ${\rm (CH_2)_2}$ \\
\hline
$\Delta g_{11}$ &&   63 &   56 &   51 &&   63 &   59 &   51 &&
                     27 &   58 &   86 \\
$\Delta g_{22}$ && 8922 & 8905 & 8882 && 8922 & 8889 & 8838 &&
                   7289 & 7278 & 7245  \\
$\Delta g_{33}$ && 4206 & 4203 & 4206 && 4206 & 4203 & 4206 &&
                   3945 & 3989 & 3791  \\
\hline
\hline
\end{tabular}
\vspace{0.1cm}\\
 $^a$ Electronic g-tensor RDFT-LR calculations
performed using BP86 exchange--correlation functional in Huz-II
basis set. $^b$ Electronic g-tensor shifts are in ppm.
\end{table}


\newpage
\begin{figure}[t]
\centerline{\epsfig{file=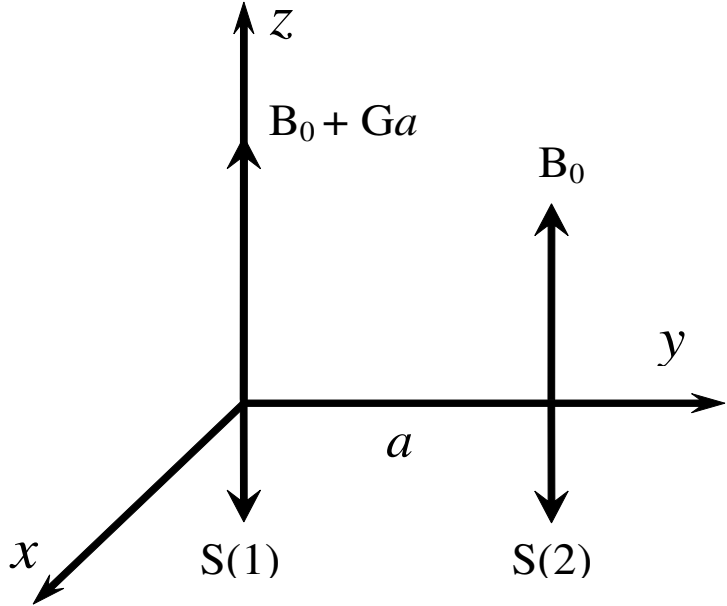,width=14cm,height=10.4cm,clip=}}
\vspace{4mm} \caption{Two spins ${\vec S}(1)$ and ${\vec S}(2)$
separated by the distance $a$. The external magnetic field ${\vec
B}(y)$ is oriented in the positive $z$-direction and has a
gradient in the $y$-direction. In the ground state, shown in the
figure, the spins point in the negative $z$-direction,
$G=|\partial B_z/\partial y|$.}
\end{figure}

\newpage
\begin{figure}[t]
\centerline{\epsfig{file=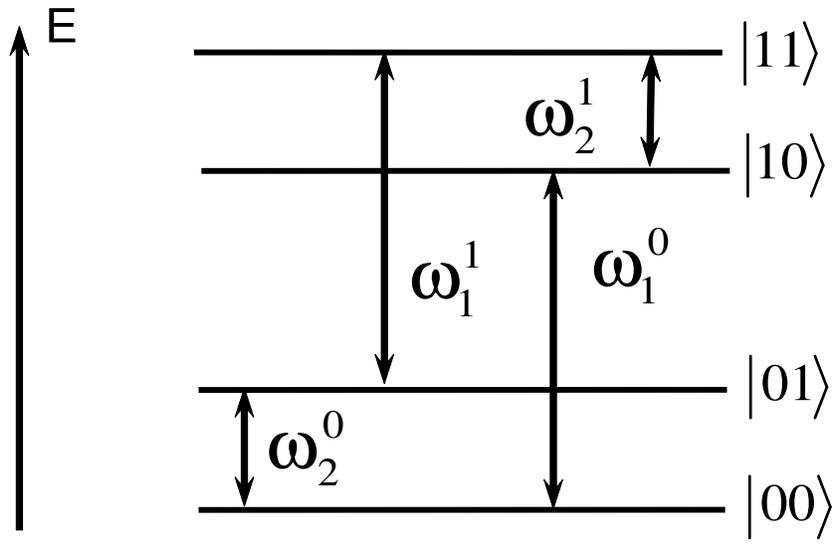,width=14cm,height=10.4cm,clip=}}
\vspace{4mm} \caption{The diagram of the energy levels for the two
qubit system.}
\end{figure}

\newpage
\begin{figure}[t]
\centerline{\epsfig{file=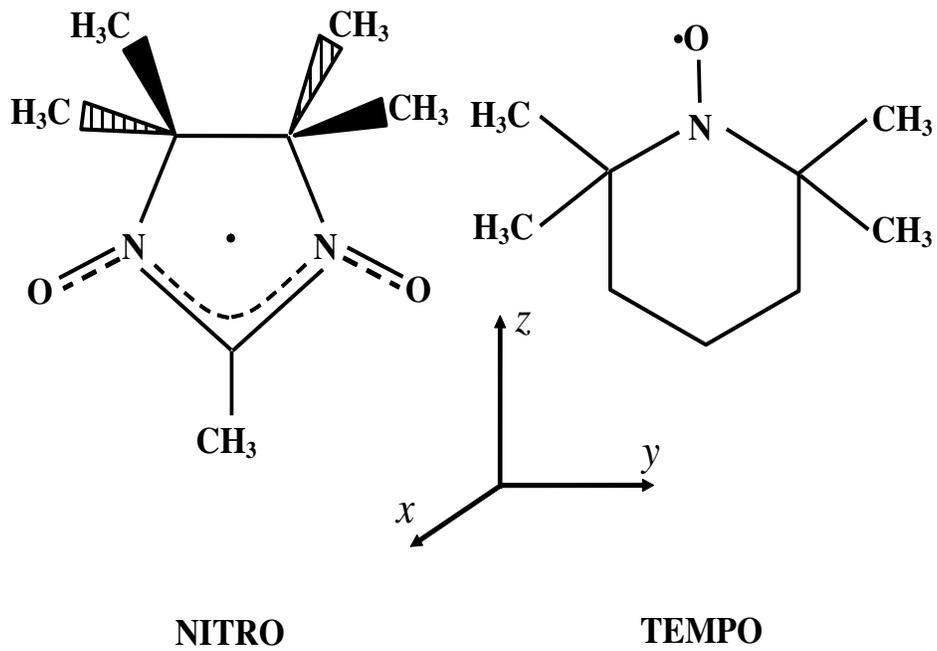,width=14cm,height=10.4cm,clip=}} \vspace{4mm}
\caption{NITRO and TEMPO radicals.}
\end{figure}

\newpage
\begin{figure}[t]
\centerline{\epsfig{file=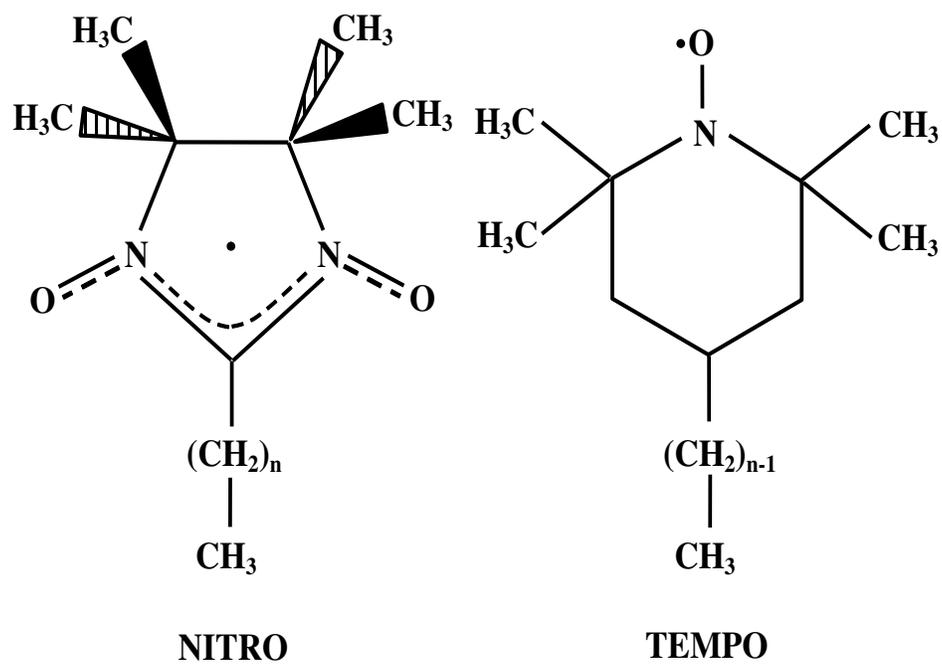,width=14cm,height=10.4cm,clip=}}
\vspace{4mm} \caption{NITRO and TEMPO radicals with $CH_2$ tails
($n=1,2$).}
\end{figure}

\newpage
\begin{figure}[t]
\centerline{\epsfig{file=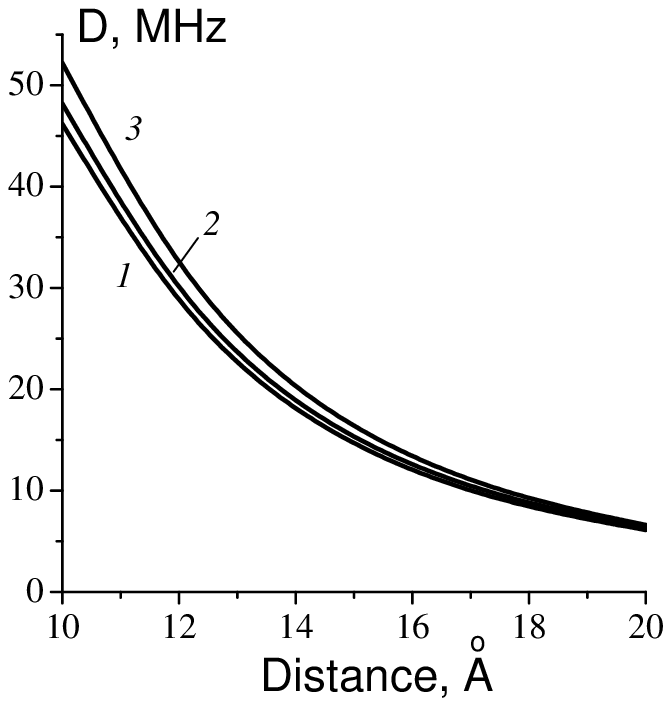,width=10cm,height=10.7cm,clip=}}
\vspace{4mm} \caption{Dependence of the parameter $D$ on the
distance between two radicals in the spin group (1 - two
interacting TEMPO radicals, 2 - two interacting NITRO radicals, 3
- two interacting point dipoles.}
\end{figure}


\begin{thebibliography}{99}
\bibitem{SCIENCE} J. Chen, M.A. Reed, A.M. Rawlett, and J.M. Tour,
SCIENCE, {\bf 286}, 1550 (1999).
\bibitem{berm02} G.~P.~Berman, V.~I.~Tsifrinovich and D.~L.~Allara, Phys. Rev. B 66, 193406 (2002).
\bibitem{tamulis}A. Tamulis, V.I. Tsifrinovich, S. Tretiak, G.P. Berman, and D.L.
Allara, quant-ph/0307136.
\bibitem{natu04} D. Rugar, R. Budakian, H.J. Mamin, and B.W. Chui,
NATURE, {\bf 430}, 329 (2004).
\bibitem{berm03} G.P. Berman, F. Borgonovi, and V.I. Tsifrinovich,
Quantum Information and Computation, {\bf 4}, 102 (2004).
\bibitem{rink03} Z.~Rinkevicius, L.~Telyatnyk, P.~Sa{\l}ek, O.~Vahtras
and H.~{\AA}gren, J. \ Chem. \ Phys. {\bf 119}, 10489 (2003).
\bibitem{malk02} O.~L.~Malkina, J.~Vaara, B.~Schimmelpfennig, M.~Munzarov{\'a}, V.~G.~Malkin and
M.~Kaupp, J.\ Am.\ Chem.\ Soc.\ {\bf 122}, 9206 (2000).
\bibitem{nees01} F.~Neese, J.\ Chem.\ Phys.\ {\bf 115}, 11080 (2001)
\bibitem{kaup02} M.~Kaupp, R.~Reviakine, O.~L.~Malkina, A.~Arbuznikov, B.~Schimmelpfenning and
V.~G.~Malkin, J.\ Computational\ Chem.\ {\bf 23}, 794  (2002).
\bibitem{dalton} T.~Helgaker, H.~J.~Aa.~Jensen, P.~J{\o}rgensen,
J.~Olsen, K.~Ruud, H.~{\AA}gren, A.~A.~Auer, K.~L.~Bak, V.~Bakken,
O.~Christiansen, S.~Coriani, P.~Dahle, E.~K.~Dalskov, T.~Enevoldsen,
B.~Fernandez, C.~H{\"a}ttig, K.~Hald, A.~Halkier, H.~Heiberg,
H.~Hettema, D.~Jonsson, S.~Kirpekar, R.~Kobayashi, H.~Koch,
K.~V.~Mikkelsen, P.~Norman, M.~J.~Packer, T.~B.~Pedersen, T.~A.~Ruden,
A.~Sanchez, T.~Saue, S.~P.~A.~Sauer, B.~Schimmelpfennig,
K.~O.~Sylvester-Hvid, P.~R.~Taylor, and O.~Vahtras, {\em DALTON, a
molecular electronic structure program, Release 1.2 (2001)}. See {\tt
http://www.kjemi.uio.no/software/dalton/dalton.html}.
\bibitem{amfi} B.~A.~Hess, C.~M.~Marian, U.~Wahlgren and O.~Gropen,
Chem.\ Phys. \ Lett. {\bf 251}, 365 (1996).
\bibitem{6311g} R.~Krishnan, J.~S.~Binkley, R.~Seeger and  J.~A.~Pople,
J. Chem. Phys. {\bf 72}, 650 (1980).
\bibitem{becke} A.D.~Becke, Phys. \ Rev.\ A {\bf 38}, 3098 (1988).
\bibitem{lyp} C.~Lee, W.~Yang and R.G.~Parr, Phys.\ Rev.\ B, {\bf 37}, 785 (1988).
\bibitem{b3hyb} A.~D.~Becke, J. \ Chem. \ Phys. {\bf 98}, 5648 (1993).
\bibitem{gaus98} M.~J.~Frisch, G.~W.~Trucks, H.~B.~Schlegel,
G.~E.~Scuseria, M.~A.~Robb, J.~R.~Cheeseman, V.~G.~Zakrzewski,
J.~A.~Montgomery, Jr., R.~E.~Stratmann, J.~C.~Burant, S.~Dapprich,
J.~M.~Millam, A.~D.~Daniels, K.~N.~Kudin, M.~C.~Strain, O.~Farkas,
J.~Tomasi, V.~Barone, M.~Cossi, R.~Cammi, B.~Mennucci, C.~Pomelli,
C.~Adamo, S.~Clifford, J.~Ochterski, G.~A.~Petersson, P.~Y.~Ayala,
Q.~Cui, K.~Morokuma, D.~K.~Malick, A.~D.~Rabuck, K.~Raghavachari,
J.~B.~Foresman, J.~Cioslowski, J.~V.~Ortiz, A.~G.~Baboul,
B.~B.~Stefanov, G.~Liu, A.~Liashenko, P.~Piskorz, I.~Komaromi,
R.~Gomperts, R.~L.~Martin, D.~J.~Fox, T.~Keith, M.~A.~Al-Laham,
C.~Y.~Peng, A.~Nanayakkara, C.~Gonzalez, M.~Challacombe,
P.~M.~W.~Gill, B.~Johnson, W.~Chen, M.~W.~Wong, J.~L.~Andres,
C.~Gonzalez, M.~Head-Gordon, E.~S.~Replogle, and J.~A.~Pople, {\em
Gaussian 98, Revision A.7\/} (Gaussian, Inc., Pittsburgh PA, 1998).
\bibitem{pw86} J.~P.~Perdew, Phys. \ Rev. \  B {\bf 33}, 8822 (1986).
\bibitem{iglo2} W.~Kutzelnigg, U.~Fleischer, and M.~Schindler, in {\em
NMR Basic Principles and Progress}, Vol.~23, edited by P.~Diehl, E.~Fluck,
H.~G\"unther, R.~Kosfeld, and J.~Seelig (Springer, Heidelberg, 1990), p.~165.
\bibitem{dz} T. H. Dunning, Jr., J.Chem.Phys. {\bf 53}, 2823 (1970).
\end{thebibliography}
\end{document}